\documentclass{iaus}
\usepackage{graphicx}

\title[H\,II region spectra to probe the ionizing radiation from massive stars] 
{Using H\,II region spectra to probe the ionizing radiation from massive stars}

\author[S. Sim\'on-D\'iaz, J. Garc\'ia-Rojas, G. Stasi\'nska \& C. Esteban]   
{S. Sim\'on-D\'iaz$^{1,3}$
  J. Garc\'ia-Rojas$^{2}$, G. Stasi\'nska$^{3}$
  \and C. Esteban$^4$}

\affiliation{$^1$Observatoire de Gen\`eve,  
 $^2$Universidad Nacional Aut\'onoma de M\'exico \\
[\affilskip] $^3$LUTH, Observatoire de Paris, Site de Meudon, 
$^4$Instituto de Astrof\'isica de Canarias
}

\pubyear{2007}
\volume{250}  
\pagerange{119--126}
\jname{Massive Stars as Cosmic Engines}
\editors{}

\newcommand{\fastwind} {{\sc fastwind}}
\newcommand{\cmfgen} {{\sc cmfgen}}
\newcommand{\tlusty} {{\sc tlusty}}
\newcommand{\wmbasic} {{\sc wm}-{\em basic}}
\newcommand{\cloudy} {{\sc cloudy}}

\newcommand{\nlte} {{\sc nlte}}
\newcommand{\lte} {{\sc lte}}
\newcommand{\hii} {H\,{\sc ii}}
\newcommand{\ion}[2] {#1\,{\sc #2}}
\newcommand{\qion}[2]{$Q$(#1\,$^{#2}$)}

\newcommand{\Te} {$T_{\rm e}$}
\newcommand{\Ne} {$N_{\rm e}$}

\begin{document}

\maketitle

\begin{abstract}
We present some results of an on-going project aimed at studying a sample of Galactic \hii\ 
regions  ionized by a single massive star to test the predictions of modern generation 
stellar atmosphere codes in the H Lyman continuum. The observations collected for this 
study comprise the optical spectra of the corresponding ionizing stars, along with imaging 
and long-slit spatially resolved nebular observations. The analysis of the stellar spectra 
allows to obtain the stellar parameters of the ionizing star, while the nebular observations 
provide constraints on the nebular abundances and gas distribution. All this information is 
then used to construct tailored photoionization models of the \hii\ regions. The reliability 
of the stellar ionizing fluxes is hence tested by comparing the photoionization model 
results with the observations in terms of the spatial variation across the nebula of an 
appropriate set of nebular line ratios.  
\end{abstract}

\firstsection 

\section{Motivation}

In the last decades, a great effort has been devoted to the development of 
stellar atmosphere codes for massive stars. The advent of the new generation 
of \nlte, line blanketed model atmosphere codes, either plane-parallel 
(\tlusty, \cite[Hubeny \& Lanz, 1995]{Hub95}), or spherically expanded 
(\fastwind, \cite[Puls et al. 2005]{Pul05}; CoSTAR, \cite[Schaerer \& de Koter 1997]{Sch97}; 
\cmfgen, \cite[Hillier \& Miller 1998]{Hil98}; \wmbasic, \cite[Pauldrach et al. 2001]{Pau01}) 
is already a fact. Each of them uses specific approximations in the calculation 
of the stellar atmosphere structure and spectral energy distribution (SED).  

This new generation of stellar atmosphere models produce quite different ionizing SEDs from the ones 
produces by the previous plane-parallel, hydrostatic models (either \lte\ by \cite[Kurucz 1991]{Kur91}, 
or \nlte\ by 
e.g. \cite[Mihalas \& Auer 1970]{Mih70}). Some notes on this, and on the consequences on the 
ionization structure 
of \hii\ regions, can be found in \cite[Gabler et al. (1989)]{Gab98}, \cite[Najarro et al. (1996)]{Naj96}, 
\cite[Sellmaier et al. (1996)]{Sel96}, \cite[Rubin et al. (1995)]{Rub95}, and 
\cite[Stasi\'nska \& Schaerer (1997)]{Sta97}. 
Although the new predictions seem to go in the 
right direction (viz. \cite[Giveon et al. 2002]{Giv02}, \cite[Morisset et al. 2004]{Mor04}) non-negligible 
differences can still be found 
between the various stellar codes (see e.g. 
\cite[Mokiem et al. 2004]{Mok04}, \cite[Martins et al. 2005]{Mar05}, \cite[Puls et al. 2005]{Pul05}). 

Since the FUV range of the stellar flux cannot be observed directly, it is crucial to find indirect tests to 
constrain it. Ionized nebulae have many times been claimed as potential tools to check the validity of the 
emergent SED predicted by the stellar atmosphere models. Here we consider the possibility of using 
tailored models of simple, spatially resolved Galactic \hii\ regions ionized by single massive stars 
to this aim. In this work we present preliminary results for two apparently spherical Galactic \hii\ regions 
(M\,43 and Sh\,2-112) to illustrate the capabilities of the methodology we have developed.

\section{Nebular candidates: the Galactic HII regions M 43 and Sh2-112}

\begin{figure}[!h]
\begin{center}
 \includegraphics[width=13 cm]{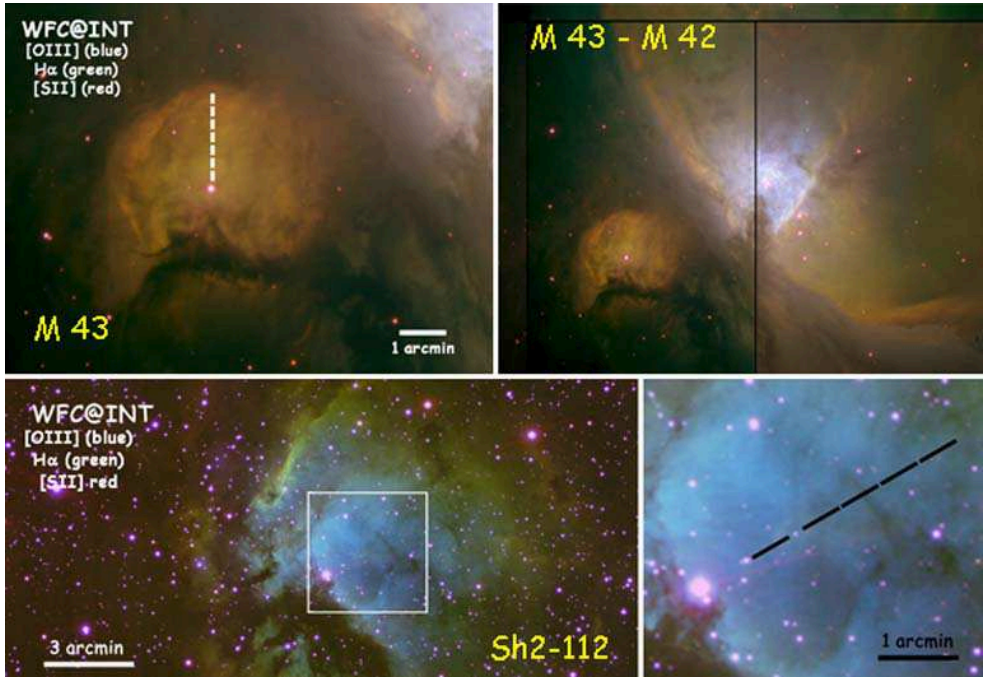} 
 \caption{WFC images of the Galactic HII regions M\,43 (up) and Sh\,2-112 (down), 
showing the location of the small apertures obtained from the long-slit 
used for the nebular spectroscopy. The ionizing stars of these nebulae are
HD\,37061 (B0.5\,V) and BD\,+35\,4216 (O8\,V), respectively.
Note the proximity of M\,43 to the Orion Nebula (M\,42) in the upper right
image.}
   \label{fig1}
\end{center}
\end{figure}

\section{The observational dataset}

\begin{figure}[!h]
\begin{center}
 \includegraphics[width=11 cm]{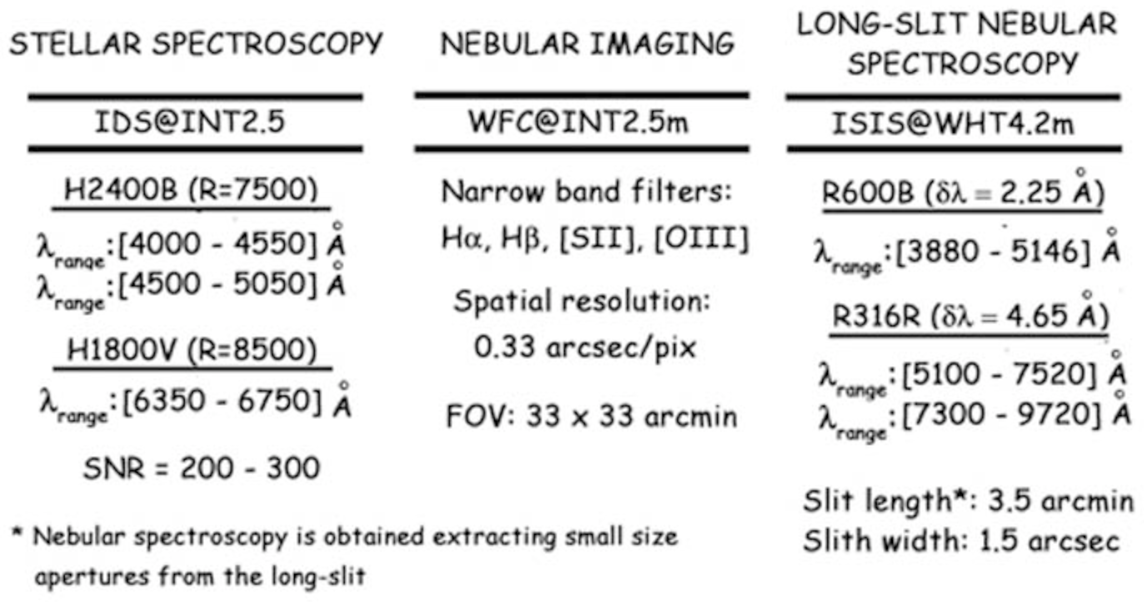} 
\end{center}
 {\bf Table 1:} Summary of the observational dataset characteristics
\end{figure}

\newpage

\section{Detailed spectroscopic analysis of the ionizing stars: HD 37061 and BD+45 3216}

\begin{figure}[!h]
\begin{center}
 \includegraphics[width=10 cm]{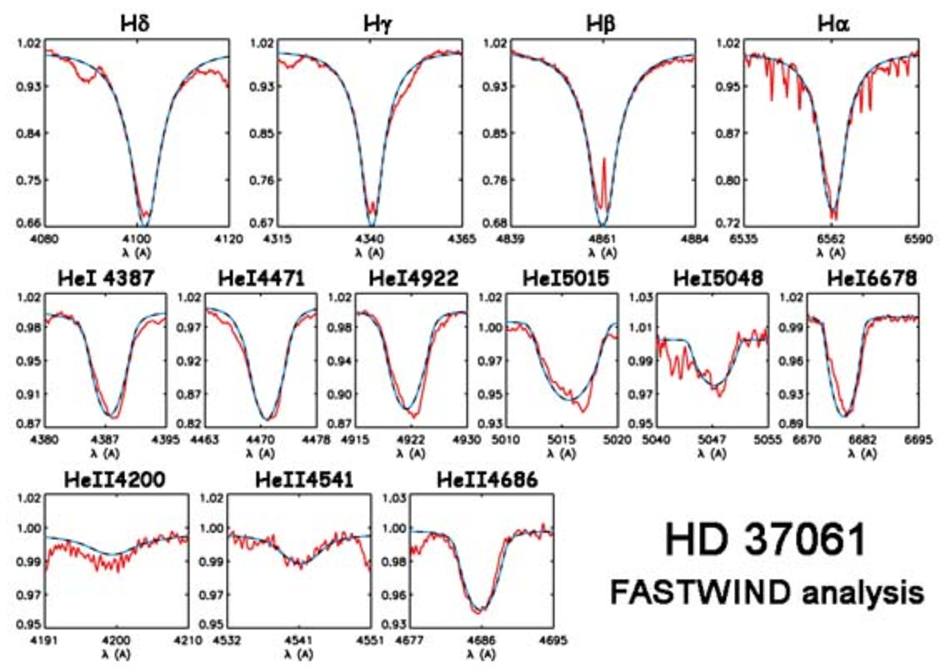} 
 \includegraphics[width=10 cm]{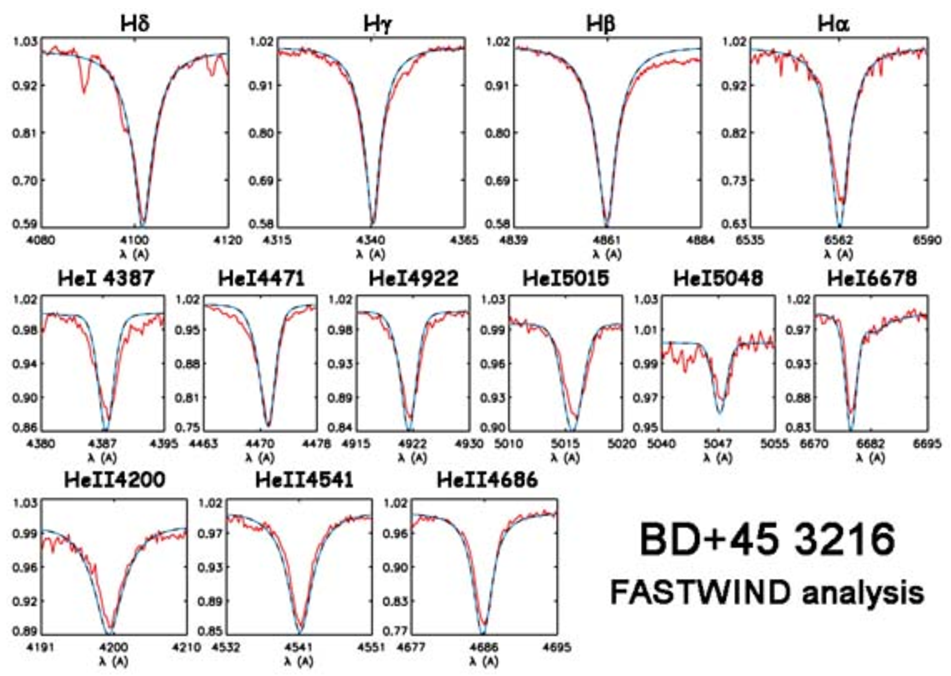} 
 \caption{The stellar parameters of the ionizing stars were determined 
 by means of the fit of the \ion{H}{}, \ion{He}{i-ii} observed profiles with \fastwind\ 
 synthetic profiles convolved with the instrumental and rotational
 profiles (see Tables 1, 2a-b). In the case of BD\,+35\,4216 an 
 extra-broadening (i.e. macroturbulence) was needed to properly fit the
 observed lines.}
   \label{fig2}
\end{center}
\end{figure}

\begin{figure}[!h]
\begin{center}
 \includegraphics[width=6.5 cm]{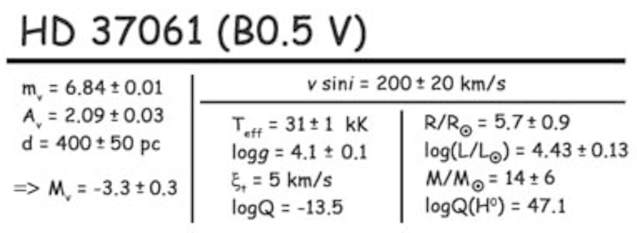} 
 \includegraphics[width=6.5 cm]{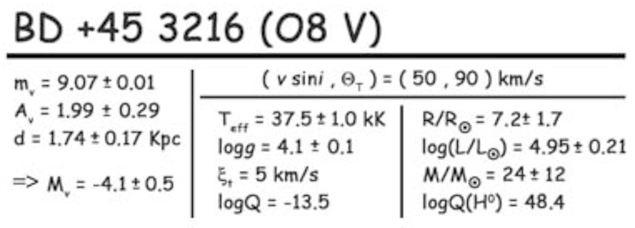} 
\end{center}
 {\bf Tables 2a-b:} Results from the spectroscopic analysis of the stars. Photometric 
 data and distances to the stars, used to estimate their stellar radii,
 luminosities and masses, are also indicated.
\end{figure}

\section{Comparing the ionizing SEDs predicted by the various stellar atmosphere codes}

As a first step, we compared the predictions --- in 
terms of the shape of the ionizing SEDs ---  of  the
four modern stellar atmosphere codes \cmfgen, 
\wmbasic, \fastwind, and \tlusty. To this aim 
we specifically calculated models with the various 
stellar codes, using  the same stellar parameters 
(those indicated in Tables 2a-b), and the solar set 
of abundances from \cite[Grevesse \& Sauval (1998)]{Gre98}. 

\begin{figure}[!h]
\begin{center}
 \includegraphics[width=6.5 cm]{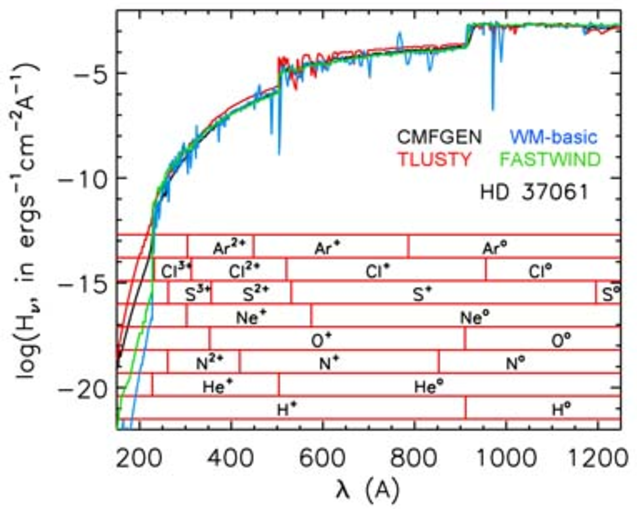} 
 \includegraphics[width=6.5 cm]{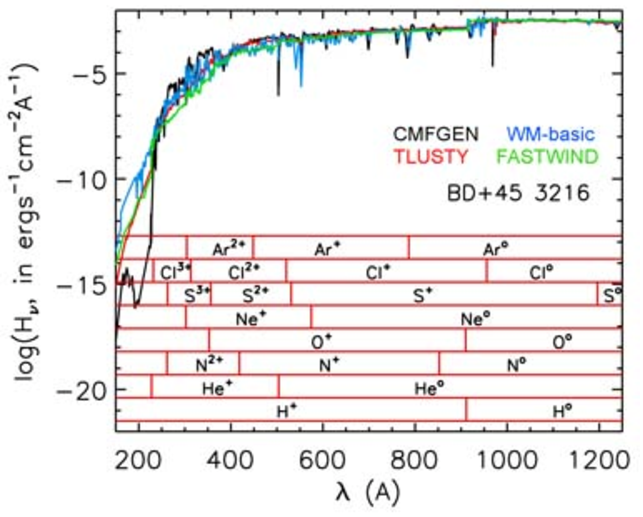} 
 \caption{Comparison of ionizing SEDs  from 
\cmfgen, \wmbasic, \fastwind, and \tlusty\ 
models with the stellar parameters indicated in 
Tables 2a-b. All SEDs have been re-mapped to the 
same $\lambda$-grid for a better comparison.}
   \label{fig3}
\end{center}
\end{figure}

\begin{figure}[!h]
\begin{center}
 \includegraphics[width=6.5 cm]{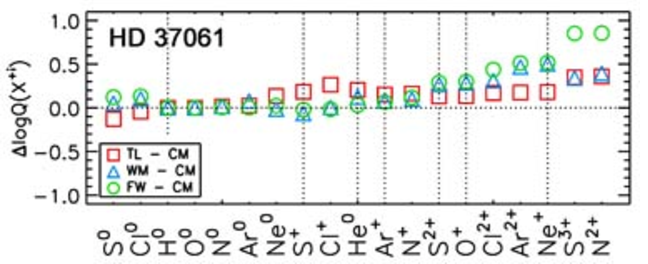} 
 \includegraphics[width=6.5 cm]{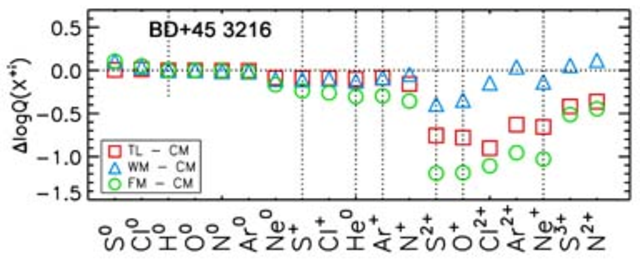} 
 \caption{To quantify the differences showed by Figs. 3, we plot the quantity 
 $\Delta$log\,\qion{X}{+i} for the 
ions indicated in those figures, using a scale in which $\Delta$log\,\qion{H}{0}=0. 
This type of diagram gives us an 
idea of how the shape of the SEDs compares in the H Lyman continuum. Alternatively, the number 
of ionizing photons for the various ionic species relative to the total number of \ion{H}{$^0$}
ionizing photons can be compared (see its utility in Sect. 6). }
   \label{fig4}
\end{center}
\end{figure}

There are several ingredients which can affect the ionizing SED predictions:\\

\begin{enumerate}
\item{The ionization structure of the stellar atmosphere}
\item{The blocking of emergent flux by the large amount of weak metal lines present in this 
     spectral region (line blocking) and its effect over the ionization structure of the stellar 
     atmosphere (line blanketing)}
\item{The presence of strong resonant lines (either in absorption or emission)}
\item{The presence of a velocity field in the stellar atmosphere}
\end{enumerate}
\vspace{0.5cm}
All these effects must be taken into account when comparing the ionizing SED. Depending 
of the spectral region, the importance of each one of them can be more or less important.

\section{The stellar - nebular connexion}

The state of ionization of a nebula depends, in
first approximation, on the ionization parameter
($U$), and the hardness of the ionizing radiation.
Following \cite[Vilchez \& Pagel (1988)]{Vil88}:

\begin{equation}
\frac{n({\rm X}\,^{i+1})}{n({\rm X}\,^{i})} \propto U 
     \frac{\int_{\nu(X^i)}^{\infty}\frac{H_\nu}{h\nu}d\nu}
	      {\int_{13.6 eV}^{\infty}\frac{H_\nu}{h\nu}d\nu} \propto U
		  \frac{{\rm Q(X}\,^{i})}{{\rm Q(H}\,^{0})},
\end{equation}

\vspace{0.5cm}

where $n$(X\,$^{i+1}$) is the number density of the i-times
ionized atom of element X, and $H_\nu$ is the stellar
Eddinton flux. Therefore, a ratio $n$(X\,$^{i+1}$)\,/\,$n$(X\,$^{i}$) 
is, to a first approach and for a given $U$,
proportional to the number of photons able to 
ionize X\,$^{i}$ relative to that of the Lyman
continuum photons.

Therefore, we can use nebular line ratios 
involving lines from to successive ions of the
same element (i.e. giving the ionization degree
of the nebula) as constraints of the ionizing
stellar SEDs at various wavelengths. These
are indicated in Table 3.                              

\section{Some useful nebular line ratios}

\begin{figure}[!h]
\begin{center}
\begin{minipage}{6.5cm}
 \includegraphics[width=6.5 cm]{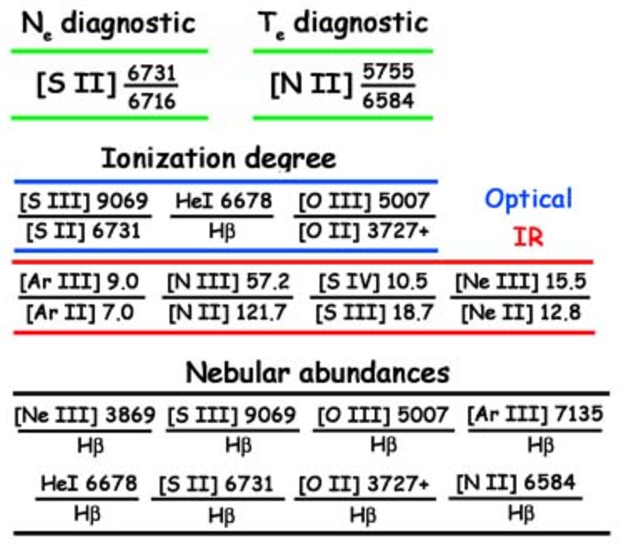} 
\end{minipage}
\   \
\hfill \begin{minipage}{6.5cm}
{\bf Table 3:} List of some useful nebular line ratios indicating the nebular 
physical properties (\Te, \Ne, and ionization degree) and 
abundances. The color code is the same as the 
one used in Figs 5 and 6.
\end{minipage}
\end{center}
\end{figure}

\section{Tailored photoionization models: probing the ionizing SED predictions}

We then constructed photoionization models for the nebulae, using 
the ionizing SEDs commented in Sec. 5. We used \cloudy\ (v02.07.01, 
\cite[Ferland et al. 1998]{Fer98}). Tables 4a-b summarize 
the characteristics of the \cloudy\ models. The nebular abundances 
were initially derived by means of the usual  nebular techniques 
(i.e. direct method), and then fine-tuned to fit the observational 
constraints.   \\

Figs. 5 and 6 compare the photoionization model results (as a function 
of the projected distance to the central star) with the nebular 
observational data obtained from the various apertures. 

\newpage

\begin{figure}[!h]
\begin{center}
 \includegraphics[width=6.5 cm]{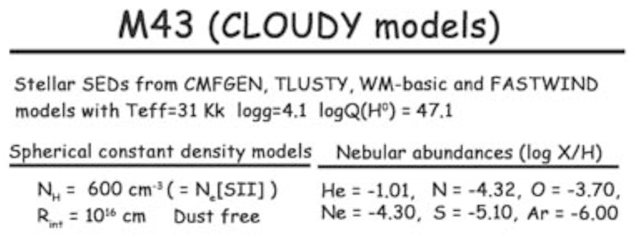} 
 \includegraphics[width=6.5 cm]{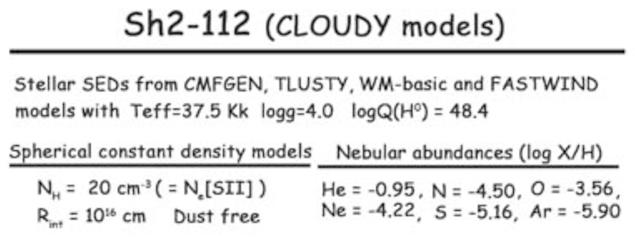} 
 {\bf Tables 4a-b:} Input parameters used for the
CLOUDY photoionization models.
\end{center}
\end{figure}

\begin{itemize}
\item[\bf{M43: }]{The output from photoionization models using SEDs from the 
various stellar codes are very similar, except for the [\ion{S}{iii}]/[\ion{S}{ii}] 
line ratio (as expected from the comparison of  the quantity \qion{S}{+}\,/\,\qion{H}{0} 
in Fig. 4). The observational constraints point towards predictions from the 
spherical models; unfortunately, the nebular spectrum of M\,43 is contaminated by an 
external nebular emission (as can be noticed from the behavior of the 
[\ion{O}{iii}]/H$_{\alpha}$, [\ion{Ne}{iii}]/H$_{\alpha}$, and [\ion{O}{iii}]/[\ion{O}{ii}]
line ratios in the outer part of the nebula), not allowing to achieve a firm conclusion.
The behavior of these line ratios may be explained by assuming that the nearby 
O7\,V star $\theta^1$\,Ori\,C (the ionizing source of M\,42; see Fig. 1) is ionizing the 
nebular material in front of M\,43. This contamination is clearly noticed when considering
lines from \ion{O}{$^{2+}$} and \ion{Ne}{$^{2+}$}, since these lines are expected to be
emitted by M\,43 itself only in a small region close to the ionizing star (due to the 
low temperature of HD\,37061), but other lines may also be affected.}
\begin{figure}[!h]
\begin{center}
 \includegraphics[width=11 cm]{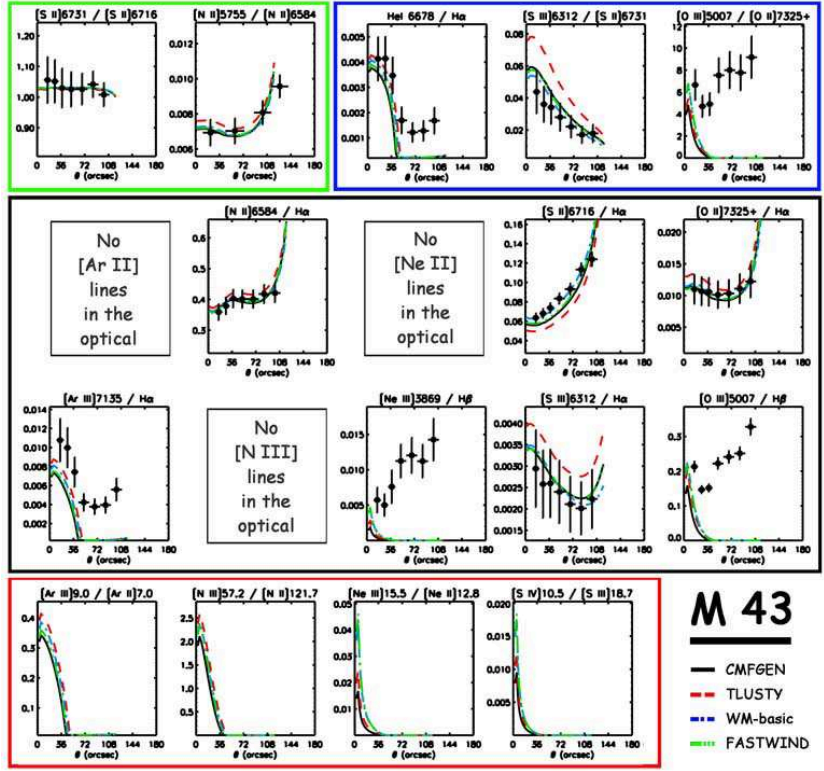} 
 \caption{Comparison of the output from 
the photoionization models with the nebular 
spectroscopic observations for the case of M43.}
   \label{fig1}
\end{center}
\end{figure}
\newpage
\item[\bf{Sh2-112: }]{In this case, photoionization models show that nebular 
results are very different depending on the considered SED. It is interesting 
to see how results presented in Fig. 6 (concerning the nebular ionization degree 
constraints) and Fig. 4 are related. Note the importance of the IR line ratios 
in this case (except for the optical [\ion{O}{iii}]\,/\,[\ion{O}{ii}] line 
ratio, it is the only way to test  the stellar SEDs for $\lambda$\,$\le$\,430\,\AA). }
\begin{figure}[!h]
\begin{center}
 \includegraphics[width=11 cm]{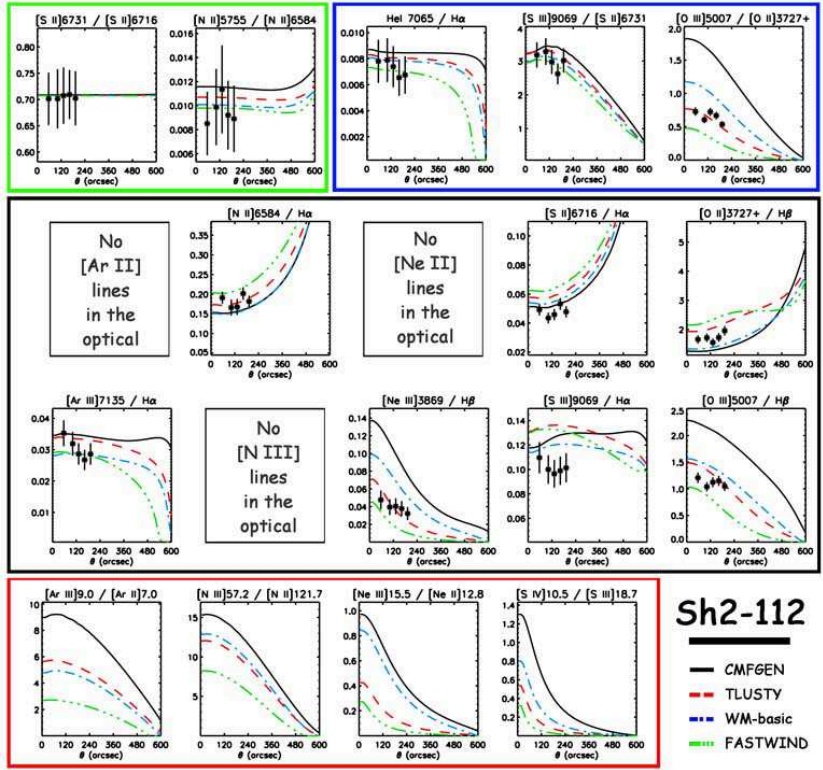} 
 \caption{As Fig. but for the case of Sh2-112.}
   \label{fig1}
\end{center}
\end{figure}
\end{itemize}

\newpage

\section{A final note}

\begin{itemize}
\item{We have shown that the methodology presented here 
(based on a combined stellar-nebular study of spatially 
resolved HII regions ionized by a single massive star) is very 
promising regarding the study of the reliability of the ionizing 
SEDs predicted by stellar atmosphere codes.}
\item{We have presented here results using simple dust free 
constant density spherical models representing the nebular 
gas, but we are also exploring the effect of considering dust, 
more complicated nebular gas distributions, and the possible 
escape of  ionizing photons in the models. It is important to 
test the effect of these ingredients before proposing 
conclusions about the ionizing SED predictions. }
\end{itemize}

\end{document}